\documentclass[12pt,a4paper]{article}
\usepackage[utf8]{inputenc}
\usepackage{amsmath}
\usepackage{amsfonts}
\usepackage{amssymb}

\title{\bf The Higgs vacuum is unstable}

\vspace{5cm}

\author{\bf Archil Kobakhidze and Alexander Spencer-Smith \\ \\
{\it ARC Centre of Excellence for Particle Physics at the Terascale, } \\
{\it School of Physics, The University of Sydney, NSW 2006 } \\
{\it E-mails: archilk@physics.usyd.edu.au, alexss@physics.usyd.edu.au}
}
\date{}

\begin{document}

\maketitle
\begin{center}

\end{center}

\begin{abstract}
\noindent
So far, the experiments at the Large Hadron Collider (LHC) have shown no sign of new physics beyond the Standard Model. Assuming the Standard Model is correct at presently available energies, we can accurately extrapolate the theory to higher energies in order to verify its validity. Here we report the results of new high precision calculations which show that absolute stability of the  Higgs vacuum state is now excluded.  Combining these new results with the recent observation of primordial gravitational waves by the BICEP Collaboration, we find that the Higgs vacuum state would have quickly decayed during cosmic inflation, leading to a catastrophic collapse of the universe into a black hole. 
Thus, we are driven to the conclusion that there must be some new physics beyond the Standard Model at energies below the instability scale $\Lambda_I \sim 10^{9}$ GeV, which is responsible for the stabilisation of the Higgs vacuum.
\end{abstract}

\baselineskip=16pt

\section{Introduction}

The Higgs mechanism \cite{Higgs:1964ia, Higgs:1964pj, Englert:1964et, Guralnik:1964eu} employed within the Standard Model provides masses for all the elementary particles, except neutrinos. These masses ($m_i$) are universally proportional to the respective constants of interaction between the elementary particles and the Higgs boson: $m_i=\lambda_i v$, where $v\equiv \left(\sqrt{2} G_F\right)^{-1/2}\approx 246$ GeV is the Higgs vacuum expectation value, which, in turn, is defined through the measurements of the Fermi constant $G_F$ in weak processes (e.g., week decays of muon). Therefore, by measuring elementary particle masses we also determine the strength of their interactions with the Higgs boson, and in particular we may measure the Higgs self-interaction constant. 

Since the current LHC Higgs data is consistent with Standard Model predictions (within the, albeit still large, error bars) \cite{Aad:2012tfa, Chatrchyan:2012ufa, Chatrchyan:2012jja, Chatrchyan:2013lba, Aad:2013wqa,  Aad:2013xqa},  and no sign of new physics has been found so far \cite{Chatrchyan:2013fea, ATLAS-newphysics}, it is reasonable to assume that the Standard Model correctly describes nature at presently probed energies. If so, with the discovery of the Higgs boson \cite{Aad:2012tfa, Chatrchyan:2012ufa} with a mass of $M_h=125.9\pm 0.4$ GeV  \cite{Beringer:1900zz} all the parameters of the Standard Model are now known. Thus, using the renormalisation group technique,  one can extrapolate the theory to higher energies to verify it's consistency. 

The previously most accurate calculations \cite{Bezrukov:2012sa, Degrassi:2012ry, Buttazzo:2013uya}  indicate that  the Higgs self-interaction coupling becomes negative at energies  $\Lambda_I \sim 10^{11}$ GeV, signalling  that the Higgs potential develops a new global minimum at large Higgs field values, $\langle h \rangle \sim 10^{17}~{\rm GeV}~ >>v$. This new minimum corresponds to the true vacuum state of the theory, which carries large negative energy density $\sim -10^{-66}$ GeV$^{4}$. A universe with a vacuum energy of that magnitude is obviously inhabitable. However, the previous calculations were not accurate enough to  establish the fate of the electroweak vacuum conclusively. 

In what follows we describe the results of a more accurate calculation which definitively establishes the fact that the electroweak Higgs vacuum, within the Standard Model, is not the true vacuum state of the theory. Then, we argue that the recent observation of primordial tensor fluctuations by the BICEP Collaboration \cite{Ade:2014xna} implies that the Higgs vacuum is actually unstable. This represents compelling evidence in favour of new physics beyond the Standard Model, new physics which must stabilise the electroweak vacuum.

\section{Higgs vacuum stability analysis}

Previously, the extrapolation of Standard Model parameters to high energies was performed by solving three-loop renormalisation group equations (RGEs) in the mass-independent modified minimal subtraction scheme ($\overline{\rm MS}$) \cite{Bezrukov:2012sa, Degrassi:2012ry, Buttazzo:2013uya}. The RGE $\beta$-functions in this scheme do not depend on particle masses and thus are simpler. The price for this simplicity, however, is that one needs to take special care at particle mass thresholds. In Ref. \cite{Buttazzo:2013uya}, the $\overline{\rm MS}$  Higgs quartic self-interaction coupling $\lambda(\mu)$, the Higgs-top Yukawa coupling $y(\mu)$ and the electroweak gauge couplings were computed by matching them with physical observables: the pole masses of the Higgs boson ($M_h$), the top quark ($M_t$), the $Z$-boson ($M_Z$), the $W$-boson ($M_W$), the $\overline{\rm MS}$ strong gauge coupling at $Z$-pole ($\alpha_3(M_Z)$), and the Fermi constant $G_F$. This matching takes place at the top quark mass threshold with 2-loop threshold corrections included.
Within this approximation it was found that stability of the  electroweak vacuum requires 
\begin{eqnarray}
M_t<  171.53\pm 0.23_{\alpha_3}\pm 0.15_{{\rm M}_h}\pm 0.15_{\rm th}~{\rm GeV}~,
\label{1}
\end{eqnarray}
where the errors in the first equation 
are due to the experimental uncertainty in the measurements of the strong coupling constant and the Higgs mass, and the theoretical uncertainty, respectively.  This bound should be compared to the recently obtained world average value for the top quark mass\cite{ATLAS:2014wva}:
\begin{eqnarray}
M_t =  173.34\pm 0.76 \pm 0.3_{\rm QCD}~ {\rm GeV} 
=  173.34\pm 0.82~{\rm GeV}~,
\label{2}
\end{eqnarray}
where the last error in the first equation reflects the uncertainty from non-perturbative QCD effects. In the second equation we have combined theoretical and experimental uncertainties in quadrature. Because of the experimental uncertainties in the Standard Model parameters (mostly in the top quark mass, Eq. (\ref{2})), and the accuracy of the previous calculations, it was not possible to establish the fate of the electroweak vacuum conclusively \cite{Buttazzo:2013uya}. Indeed, the boundary value in (\ref{1}) is only $1.8 \sigma$ (i.e., 96.40\% CL one sided) away from the central experimental value in (\ref{2}). The central value of the Higgs mass is $2.2 \sigma$ away from the stability bound on the Higgs mass obtained in \cite{Buttazzo:2013uya}.   

With the goal of increasing the precision of the stability bound calculation we concentrate on a more accurate treatment of particle mass threshold effects within the framework of a mixed renormalisation scheme. Namely, we computed all one-loop $\beta$-functions necessary for the running of the Higgs self-coupling in a mass-dependent renormalisation scheme, while $\overline{MS}$ $\beta$-functions were used in the two and three-loop approximation. In computing one-loop mass dependent $\beta$-functions special care has been taken to ensure gauge invariance and a correct treatment of the imaginary parts of couplings at each particle threshold. The obtained $\beta$-functions smoothly approach their corresponding one-loop $\overline{MS}$ counterparts above and below particle mass thresholds. Within this scheme threshold and decoupling effects are accounted for exactly at the one-loop level, whilst at higher order we adopt the two-loop threshold corrections from the previous calculations \cite{Buttazzo:2013uya}. The technical details of this mixed renormalisation scheme, together with a comprehensive exposition of numerical results will be presented in separate publications \cite{Alex}. Here we report a new vacuum stability bound on the top quark mass:
\begin{equation}
M_t < 170.16\pm 0.22_{\alpha_3} \pm 0.13_{M_h}\pm 0.06_{\rm th}~ {\rm GeV}~.
\label{3}
\end{equation}
The estimate of theoretical uncertainty in the above equation was obtained by following the prescription given in \cite{Degrassi:2012ry, Buttazzo:2013uya}. The upper $1 \sigma$ value from the above bound, $M_t^{\rm max}=170.42$ GeV, is $3.6 \sigma$ away from the central experimental value in (\ref{2}). That is to say, the measured top quark mass is incompatible with absolute stability of the Higgs vacuum within the Standard Model  at 99.98\% C.L. (one sided). 

Also, within our scheme  the instability scale (defined as the scale at which the running Higgs quartic coupling vanishes) is lower than in previous calculations \cite{Buttazzo:2013uya}:
\begin{equation}
\log_{10}\frac{\Lambda_I}{\rm GeV}= 9.19 \pm 0.65_{M_t} \pm 0.19_{M_H} \pm 0.13_{\alpha_3}\pm 0.02_{\rm th} = 
9.19\pm 0.69~,
\label{4}
\end{equation}  
where the errors are summed in quadrature in the last equation. The scale at which the effective Higgs self-coupling (taken from the effective potential) vanishes is roughly an order of magnitude grater than that in \eqref{4}. The $\beta$-function for the running Higgs quartic coupling at the instability scale is negative and varies in the range $\beta_{\lambda}(\Lambda_I)\equiv \bar\beta_{\lambda}=-0.4562  \pm 0.0985$, depending on uncertainties in the input parameters. These findings have important cosmological implications as we will discuss in the following.

\section{The fate of the Higgs vacuum during inflation}

Cosmic inflation is an attractive theoretical scenario since it solves many of the problems with standard hot Big Bang cosmology. 
In recent years further experimental hints have accumulated in favour of cosmic inflation. The Planck Collaboration extracted a high precision value of the spectral index, $n_s=0.9603\pm 0.0073$, from measurements of temperature anisotropies in the cosmic microwave background radiation (CMBR), thus ruling out  the phenomenologically exact scale invariant value $n_s=1$ at over $5\sigma$ \cite{Ade:2013uln}. This deviation from exact scale invariance  is predicted in all realistic inflationary models.  Most notably, the BICEP Collaboration recently announced the discovery of B-mode polarisation in the CMBR \cite{Ade:2014xna}. These polarisation modes are widely attributed to tensor perturbations (gravitational waves) generated during inflation. From the measured tensor-to-scalar amplitude ratio  \cite{Ade:2014xna}
\begin{equation}
r=0.2^{+0.07}_{-0.05}~,
\label{5}
\end{equation}
one can infer an inflationary expansion rate: 
\begin{equation}
H_{\rm inf}\approx  10^{14}~{\rm GeV}~.
\label{6}
\end{equation}

Given such a large rate of inflation, the transition from a metastable electroweak Higgs vacuum to the true vacuum (with large negative energy density) is dominated by the Hawking-Moss instanton \cite{Hawking:1981fz}. This corresponds to a process in which the Higgs field across the the entire inflationary patch develops  a large expectation value equal to the top of the potential barrier separating false and true vacua, $h_*\approx e^{-1/4}\Lambda_I$.  Subsequently, the Higgs field quickly rolls down a steep potential hill towards the negative energy true vacuum. Effectively, the decay of the electroweak Higgs vacuum can be viewed as the thermally activated nucleation of bubbles of true vacuum \cite{Brown:2007sd}, driven by the temperature of de Sitter space $T_{\rm dS}=H_{\rm inf}/2\pi$. 

The probability that a transition from the (electroweak) false vacuum, $h=v$, to the state $h_*$ occurred during `visible' inflation,  $N_e=\tau_{\rm inf}H_{\rm inf}\approx 60$, is $\left(1-e^{-p}\right)$, where \cite{Kobakhidze:2013tn}:
\begin{equation}
p=N_e^4\exp\left\lbrace \frac{\pi^2\bar\beta_{\lambda}}{2e}\frac{\Lambda_I^4}{ H^4_{\rm inf}} \right\rbrace~.
\label{7}
\end{equation}         
Combining the low instability scale calculated above (\ref{4}), the high inflationary rate extracted from the BICEP measurements (\ref{6}), and the fact that $\bar\beta_{\lambda} < 0$, we find that $p$ in (\ref{7}) is a large positive number: $p\approx N_e^4$, and hence, the electroweak Higgs vacuum decays with almost unit probability. Upon such decay the negative energy density of the Higgs potential would dominate over the positive potential energy density of the inflaton, $V_{\rm inf}\approx H_{\rm inf}^2M_P^2\approx 10^{64}$ GeV$^4$, so that expansion would become contraction and the universe would eventually end up in a black hole. We also found that the top quark mass must be $3.4 \sigma$ away from the experimental value (\ref{2}) ($M_t< 170.54$ GeV) in order to avoid such a disastrous event.    

One could argue that some patches of universe could survive inflation within the multiverse picture of eternal inflation \cite{Guth:2007ng} and use anthropic reasoning to explain our observations of an electroweak vacuum with $v \approx 246$ GeV. However, it is easy to convince yourself that Higgs vacuum decay is fast enough to cease inflation not only within a Hubble volume but also globally. Indeed, the fraction of an initial Hubble volume that is still inflating after time $\tau$ is ${\rm e^{3H_{\rm inf}\tau}}{\rm e}^{-(\tau H_{\rm inf})^4p}$. The time at which inflation stops globally is then $\tau_{\rm stop}=\left(3/p\right)^{1/3}/H_{\rm inf}$, that is, $\approx 1.4$-Hubble time or less. Therefore, eternal inflation is not possible. 

Thus we reach an important conclusion: the Standard Model Higgs vacuum  is short-lived in a universe with high inflationary rate, as it is suggested by the resent BICEP results. It would have quickly decayed during cosmic inflation, leading to the catastrophic collapse of the universe into a black hole. This conclusion is obviously altered if some new physics affects the Higgs potential at large field values. As the simplest example, one may consider inflaton-Higgs interactions, or non-minimal Higgs-gravity coupling, which induces a large effective mass for the Higgs boson during inflation. Such interactions may suppress the mechanism of Higgs vacuum decay \cite{Lebedev:2012sy, Busoni:2014sya}  induced by Higgs superhorizon quantum fluctuations during inflation \cite{Espinosa:2007qp}. However, as has been argued in \cite{Kobakhidze:2013tn}, in this situation the dominant source of Higgs vacuum decay is the Coleman -- de Lucia type transition, which is unacceptably fast. Thus, scenarios with a large effective mass for the Higgs boson are excluded. All the simple models in which the Higgs field itself drives inflation with non-minimal Higgs-gravity couplings  \cite{Bezrukov:2007ep} are excluded as well. 

On the other hand, there are strong indications of physics beyond the Standard Model related to the observation of neutrino oscillations and dark matter.  At first glance none of these seem related to the Higgs vacuum stability problem, however, it is reasonable to contemplate that the physics behind neutrino oscillations (and thus their masses) or dark matter may also be responsible for the stabilisation of the Higgs vacuum. A thorough analysis of Higgs vacuum stability within the popular models of neutrino mass generation was performed in  \cite{Kobakhidze:2013pya} and potentially interesting scenarios have been identified. Since the new physics must enter at energy scales $\lesssim \Lambda_I\sim 10^{9}$ GeV, we may even be lucky enough to observe related new particles and interactions at the LHC or in near-future high energy experiments.    

\paragraph{Acknowledgement} This work was partially supported by the Australian Research Council.


\end{document}